# On the use of mixed potential formulation for finite-element analysis of large-scale magnetization problems with large memory demand

ALEXANDER CHERVYAKOV

Laboratory of Information Technologies, Joint Institute for Nuclear Research,
Joliot Curie 6, 141980 Dubna, Russia
*acher@jinr.ru*

Abstract

The finite-element analysis of three-dimensional magnetostatic problems in terms of magnetic vector potential has proven to be one of the most efficient tools capable of providing the excellent quality results, but becoming computationally expensive when employed to modeling of large-scale magnetization problems in the presence of applied currents and nonlinear materials due to subnational number of the model degrees of freedom. In order to achieve a similar quality of calculation at lower computational cost, we propose to use for modeling such problems the combination of magnetic vector and total scalar potentials as an alternative to magnetic vector potential formulation. The potentials are applied to conducting and nonconducting parts of the problem domain, respectively and coupled together across their common interfacing boundary. For nonconducting regions, the thin cuts are constructed to ensure their simply connectedness and therefore the consistency of the mixed formulation. The implementation in the finite-element method of both formulations is discussed in detail with difference between the two emphasized. The numerical performance of finite-element modeling in terms of combined potentials is assessed against the magnetic vector potential formulation for two magnetization models, the Helmholtz coil, and the dipole magnet. We show that mixed formulation can provide a substantial reduction in the computational cost as compared to its vector counterpart for a similar accuracy of both methods.

Introduction

FEM simulations of large-scale magnetostatics problems in 3D such as the magnetization of magnets can be computationally expensive due to complexity of geometries separated differing media, nonlinearity of involved materials and enhanced requirements for accuracy of computations [1-4]. The solution of such problems is therefore limited by the available hardware resources. In particular, the capacity of the random-access memory (RAM) is crucially affected by direct solvers despite their out-of-core mechanism of distributing memory, while the CPU-processing time is considerably elongated by iterative solvers. A common tool for FEM calculation of the magnetostatic fields in the presence of currents and nonlinear materials is based on the magnetic vector potential (MVP) formulation [5-7]. Although this approach provides a superior quality of calculations, exploiting vector potential for entire problem domain including large current-free regions increases the total number of the model degrees of freedom (DOFs) and therefore the usage of computational memory and time. In order to reduce the computational cost, the regions that are free from the currents can be modelled instead with magnetic scalar potential (MSP) while leaving the current-carrying regions for the use of magnetic vector potential and coupling both potentials on their common interfacing boundaries. However, such a combination of vector and scalar potentials when applied to conducting and nonconducting regions of the problem domain, respectively must be consistent with Ampere's law unless the nonconducting regions are made simply connected in the presence of currents. A typical example is provided by the geometry of a circular coil surrounded by large insulating region of air in which the magnetic scalar potential is initially multi-valued. There are several ways to overcome such a contradiction, the simplest being to extend the vector potential region to the hole area which is a part of the insulator. However, the most effective method discussed in this paper, consists of constructing thin cuts to prevent all paths from linking the currents and imposing across each cut surface the potential jump equal to enclosed current [8-11]. Although creating cuts may not be quite simple for nontrivial model geometries, it allows for substantial reduction in the computational cost. Moreover, for some problems the application of the total scalar potential can be extended to the whole computational domain where the exciting currents are represented as line currents flowing along the cut boundaries [12,13]. In this paper, the numerical potential of mixed formulation is evaluated for two modeling examples, the Helmholtz coil surrounded by air and the dipole magnet, and compared with the potential of the MVP formulation. We show that the use of mixed formulation results in substantial reduction of computational expenses while keeping almost the same quality of numerical computation.

1. Modeling Framework

   a. Magnetostatics Equations

In magnetostatics (see, e.g. [14]), the fields caused by steady currents are described as a static limit of Maxwell's equations given by magnetic Gauss's and Ampere's laws explicitly as

$$\nabla \times \boldsymbol{H} = \boldsymbol{j}, \quad in\ \Omega, \qquad (1)$$
$$\nabla \cdot \boldsymbol{B} = 0, \quad in\ \Omega, \qquad (2)$$

where bold letters are referred to three-component vectors, $\boldsymbol{j}$ is a divergence-free ($\nabla \cdot \boldsymbol{j} = 0$) source current density, $\boldsymbol{B}$ and $\boldsymbol{H}$ are the magnetic flux density and magnetic field strength, respectively. They are further related to each other via the constitutive relation accounting for macroscopic properties of the materials

$$\boldsymbol{B} = \mu \boldsymbol{H}, \quad in\ \Omega, \qquad (3)$$

with $\mu$ being the permeability of medium. Most of the magnetostatic problems involve several materials of different permeabilities including those with highly nonlinear behavior of the corresponding magnetization curves $\mu(H)$. For such problems, $\mu$ is continuous within a material and discontinuous across the material interfaces.

Equations (1-3) hold for the whole problem domain $\Omega$, ensuring simultaneously the continuity of the tangential component of the vector $\boldsymbol{H}$ (with no surface currents) and the normal component of the vector $\boldsymbol{B}$ on its boundary $\Gamma = \partial \Omega$:

$$\boldsymbol{n} \times \boldsymbol{H} = \boldsymbol{0}, \quad on\ \Gamma_h, \qquad (4)$$
$$\boldsymbol{n} \cdot \boldsymbol{B} = 0. \quad on\ \Gamma_b, \qquad (5)$$

where $\Gamma_h$ and $\Gamma_b$ are the two complementary parts of entire boundary $\Gamma = \Gamma_h \cup \Gamma_b$ distinguished in accordance with antisymmetry/symmetry conditions (also known as the perfect magnetic conductor/ the magnetic insulation conditions), respectively. The boundary conditions (4) and (5) are imposed to truncate/magnetically insulate the whole model geometry. They can also be used as interface conditions separating different media.

Given relation (3), the equations (1) and (2) form a first order div-curl system of four scalar equations in three unknowns. In principle, FEM analysis of such a system, e. g. in terms of the flux density $\boldsymbol{B}$, can be done by using the least square method (LSFEM) with residuals including both equations (1) and (2) and the boundary condition (5) as well. However, the LSFEM does not work well for the problems with abrupt changes in material coefficients [15]. A standard approach is therefore to formulate the problem in terms of potentials rather than in primary field variables. Depending on the problem objectives, the fields can be expressed either by the magnetic scalar potential (MSP), or by the magnetic vector potential (MVP) considered as new dependent variables. The MSP formulation obviously requires less degrees of freedom than MVP for finite element modeling, but may be mathematically more elaborated in the presence of currents. In order to reduce the total number of DOFs and increase the efficiency of computation, it is also be possible to use the combined formulation applying one of the potentials to a certain region of the problem domain. The comparative analysis of using these formulations is presented in the present paper.

   b. MSP formulation

Introducing the magnetic scalar potential $V_m$ originally begins with the regions without currents ($\boldsymbol{j} = \boldsymbol{0}$), where according to equation (1) the magnetic field $\boldsymbol{H}$ becomes purely irrotational, and therefore can be represented as a negative gradient of the scalar potential $\boldsymbol{H} = -\boldsymbol{\nabla} V_m$ to satisfy Ampere's law identically. In terms of the scalar potential, the magnetic Gauss's law (2) is then described with the help of the constitutive relation (3) by Laplace's equation explicitly as

$$\boldsymbol{\nabla} \cdot (\mu \boldsymbol{\nabla} V_m) = 0, \quad in\ \Omega, \tag{6}$$

where the permeability $\mu$ reads

$$\mu(H) = B(H)/H, \quad in\ \Omega \tag{7}$$

and $B(H)$ is a material-dependent magnetization curve. The corresponding to (4) and (5) boundary conditions take the following form

$$V_m = 0, \quad on\ \Gamma_h, \tag{8}$$

$$\boldsymbol{n} \cdot \boldsymbol{\nabla} V_m = 0, \quad on\ \Gamma_b. \tag{9}$$

The zero magnetic scalar potential boundary condition (8) (or, the antisymmetry condition) enforces the continuity of scalar potential across the interface of different media preserving simultaneously the uniqueness of its definition. Indeed, although potential can be uniquely defined at one reference point chosen arbitrarily somewhere in any part of the problem domain, this choice should be common for all other adjacent parts of such domain, that is along the common boundary interface what is exactly stated by equation (8), where the zero scalar potential point is specified.

When currents are involved in the problem domain, the scalar potential becomes undefined in conducting regions and multivalued in nonconducting regions if they are multiply connected [9-11]. One way to cope with the second problem is to construct thin cuts preventing any closed path from linking the currents and impose the potential discontinuity across different sides of each cut surface as follows

$$\Delta V_m = V_m^+ - V_m^- = I, \quad on\ \Gamma_{cut}. \tag{10}$$

Equation (10) ensures that the induced in nonconducting region magnetic field corresponds to exciting current flowing in conducting loop. On finite element basis, the scalar potential is approximated by Lagrange nodal elements with DOFs specified by its value at each node. These nodes are then duplicated on the cut surfaces to distinguish between the two sides of each surface and constrain the potential difference according to equation (10). In this way, the scalar potential becomes a single-valued function everywhere in nonconducting regions surrounded currents as far as they are made simply connected. The first problem is then resolved either via introducing the magnetic vector potential for conducting regions and using mixed vector-scalar formulation for the whole problem domain, where both potentials are coupled together across common interfaces, or by representing the exciting currents as line currents flowing along the cut boundaries, whose values are accounted for by the potential discontinuities, extending thereby the MSP formulation to entire problem domain.

Multiplying equation (6) by a scalar test function $\zeta$ with the boundary condition ($\zeta = 0$ on $\Gamma_h$), integrating by parts over the computational domain $\Omega$ and using boundary condition (9), we obtain the weak formulation in terms of scalar potential

$$\int_\Omega (\mu \boldsymbol{\nabla} V_m) \cdot (\boldsymbol{\nabla} \zeta) dv - \int_{\Gamma_{cut}} \mu \zeta \boldsymbol{n} \cdot (\boldsymbol{\nabla} V_m^+ - \boldsymbol{\nabla} V_m^-) ds = 0, \qquad (11)$$

where $\Gamma_{cut}$ is a cut surface, while the function $\zeta$ after discretizing on a finite-element mesh represents a set of nodal basis functions $\zeta_i$ used to approximate the scalar potential $V_m$ following to Galerkin method. In the presence of several cuts, the last term in (11) is extended to sum over all cut surfaces.

c. MVP formulation

The number of equations (1) and (2) can also be reduced by accounting for solenoidality of the magnetic flux density $\boldsymbol{B}$ and introducing the magnetic vector potential $\boldsymbol{A}$ as $\boldsymbol{B} = \boldsymbol{\nabla} \times \boldsymbol{A}$. This satisfies identically equation (2) describing Gauss's law, while equation (1) for Ampere's law takes the form of a double curl equation

$$\boldsymbol{\nabla} \times (1/\mu \, \boldsymbol{\nabla} \times \boldsymbol{A}) = \boldsymbol{j}, \qquad in \ \Omega, \qquad (12)$$

where the permeability $\mu$ is now expressed as

$$\mu(B) = B/H(B), \qquad in \ \Omega \qquad (13)$$

and $H(B)$ is a magnetization curve specified for each material. In terms of vector potential, the corresponding to (4) and (5) boundary conditions read respectively,

$$\boldsymbol{n} \times \boldsymbol{\nabla} \times \boldsymbol{A} = \boldsymbol{0}, \qquad on \ \Gamma_h, \qquad (14)$$

$$\boldsymbol{n} \times \boldsymbol{A} = \boldsymbol{0}, \qquad on \ \Gamma_b, \qquad (15)$$

where the magnetic insulation boundary condition (15) (or, the symmetry condition) enforces the tangential continuity of the potential $\boldsymbol{A}$ across the boundary and follows directly from its definition.

The vector potential is not uniquely defined unless its divergence is specified. Without gauge fixing, the numerical modeling in three dimensions would be faced with instabilities and/or singularities of the algebraic system corresponding to equation (12). In fact, the nontrivial null space of the curl-curl operator in equation (12) involving the gradient of an arbitrary scalar field appears due to gauge freedom in the definition of the magnetic vector potential [16,17]. However, despite the ambiguity of the potential, the magnetic field and the flux density are always uniquely calculated by its means. Moreover, when only these physical quantities are matters of interest, the singular algebraic system for unknown potential can be handled by iterative solvers due to their inherent auto-gauging properties.

On the other hand, the gauge fixing might be necessary when either the uniqueness of solution for unknown potential is required, or the only direct solvers are available for its numerical computation [18]. As the MVP formulation relies merely on the curl of the vector potential, its divergence can be chosen arbitrarily. A common choice corresponds to Coulomb gauge

$$\boldsymbol{\nabla} \cdot \boldsymbol{A} = 0, \qquad on\ \Omega \tag{16}$$

Once imposed, the gauge (16) would result in yet another boundary condition

$$\boldsymbol{n} \cdot \boldsymbol{A} = 0, \qquad in\ \Gamma_h \tag{17}$$

implying the normal continuity of the magnetic vector potential across the boundary. By equations (17) and (15), the gauged $\boldsymbol{A}$-field becomes continuous in both the tangential and the normal directions to the boundary surface. Although such a continuity can be preset automatically by the approximation of the potential $\boldsymbol{A}$ on the nodal finite-element basis with its three components specified at each node, this would not be done without loss of accuracy for the problems with discontinuous medium properties, where the normal component of the potential is changed abruptly at material interfaces [6]. In modeling such problems, the common approach is therefore to approximate the vector potential with only tangential component continuous while leaving its normal component free to jump across the boundary by using instead of nodal the edge (curl-conforming) elements with DOFs assigned in a finite-element mesh along each edge rather than at each node [19,20].

The edge elements are not divergence-free in general to enforce Coulomb gauge (16) automatically on finite element basis [19]. This gauge is therefore imposed in a weak form via introducing a Lagrange multiplier $\psi$ as a new dependent variable [20-23]. Applying $\psi$ to Ampere's law results in the two coupled equations, one for the multiplier $\psi$ given by the condition (16) and another one for the potential $\boldsymbol{A}$ expressed by the modified double curl equation as compared to (12) as follows

$$\boldsymbol{\nabla} \times \left( {}^{1}\!/_{\mu}\ \boldsymbol{\nabla} \times \boldsymbol{A} \right) = \boldsymbol{j} + \boldsymbol{\nabla}\psi, \qquad in\ \Omega, \tag{18}$$

where the value of variable $\psi$ is set to a constant on insulation boundaries to avoid possible singularities.

The weak formulation in terms of magnetic vector potential is based on equations (16)-(18). Simplest, the equations (16) and (17) are represented by the following integral equation

$$\int_{\Omega} \boldsymbol{A} \cdot (\boldsymbol{\nabla}\zeta) dv = 0, \tag{19}$$

where $\zeta$ is a scalar test function satisfying the boundary condition ($\zeta = 0\ on\ \Gamma_b$). The weak form of equation (18) is then obtained multiplying (18) by the vectorial test function $\boldsymbol{w}$ with the boundary condition ($\boldsymbol{n} \times \boldsymbol{w} = \boldsymbol{0}, on\ \Gamma_b$), integrating by parts over the computational domain $\Omega$ and using boundary condition (14) explicitly as

$$\int_{\Omega} ({}^{1}\!/_{\mu}\ \boldsymbol{\nabla} \times \boldsymbol{A}) \cdot (\boldsymbol{\nabla} \times \boldsymbol{w}) dv = \int_{\Omega} \boldsymbol{j} \cdot \boldsymbol{w}\ dv + \int_{\Omega} (\boldsymbol{\nabla}\psi) \cdot \boldsymbol{w} dv, \tag{20}$$

where $w$ is discretized on a finite element mesh to represent a basis of the edge functions $w_i$ used for approximation of the potential $A$ following to Galerkin method. The equations (19) and (20) form a system of two coupled integral equations possessing unique solvability [20-23]. Aside from the gauge fixing, this system uses a gradient of $\psi$ to eliminate any divergence from the externally applied current density $j$ thereby ensuring the current continuity inherent in Ampere's law. This would preserve the conservation of current in numerical calculation which is fulfilled automatically in the analytical sense but not obviously satisfied when interpolated on the finite-element functional basis.

d. Mixed formulation

A typical computational domain of the magnetostatic problems consists of the current-carrying region surrounded by the air and other nonconducting domains. Such a geometry suggests introducing the so-called mixed vector-scalar formulation solving Eqs. (1-5) for magnetic vector potential $A$ in the regions involving currents and for total scalar potential $V_m$ in the current-free regions. As will be shown later, this leads to a substantial reduction in the computational cost. In the presence of currents however, the magnetic scalar potential would only be consistently applied to current-free regions if they are simply connected meaning that no Ampere's loops enclosed the current-carrying region are allowed there. Otherwise, the line integral of the magnetic field along a path enclosing the current would be equal either to zero, since the field is a gradient of scalar potential, or to the value of the enclosed current due to Ampere's law. As the scalar potential is assumed to be a single-valued function, such paths cannot be allowed. Typical geometries in which the scalar potential regions are not simply connected are those that surround the vector potential regions possessing the holes. To make them simply connected, it is sufficient to prevent any closed path from linking a current by constructing the cut surfaces spanning the current-carrying regions. Simultaneously, this imposes the potential discontinuity across two different sides of each cut surface equal to value of the enclosed current. In this way, the scalar potential becomes a single-valued function encountering however the jumps across each cut surface. On the finite-element basis, the vector potential is approximated by edge elements with DOFs specified by its tangential component along each edge, while the scalar potential by Lagrange nodal elements with DOFs specified by its value at each node. The conversion from nodal to edge elements can be performed [24].

The two potentials should be coupled on their common interfacing boundaries. The corresponding interface conditions follow directly from equations (4) and (5) and guarantee the continuity of the tangential component of the field $H$ as well as the normal component of the flux $B$. The equation (4) is used to specify the vector potential in terms of scalar potential, while the equation (5) to specify the scalar potential in terms of vector potential, so that the resulting conditions expressed by means of these potentials at interfacing boundary between the MVP and MSP regions read

$$(1/\mu_c) \cdot \boldsymbol{n} \times (\boldsymbol{\nabla} \times \boldsymbol{A}) = - \boldsymbol{n} \times \boldsymbol{\nabla} \cdot V_m, \qquad \text{on } \Gamma_i, \tag{21}$$

$$- \mu_n \cdot \boldsymbol{n} \cdot \boldsymbol{\nabla} \cdot V_m = \boldsymbol{n} \cdot (\boldsymbol{\nabla} \times \boldsymbol{A}), \qquad \text{on } \Gamma_i, \tag{22}$$

where $\Gamma_i$ is common interfacing boundary, while subscripts $c$ and $n$ refer to conducting and nonconducting regions, respectively. Note that the normal component of the curl of $A$ in the right-hand side of equation (22) is completely defined by the tangential component of $A$.

The weak form of the mixed vector-scalar formulation is essentially based on equations (19) and (20) applied to conducting domain $\Omega_c$ as well as on equation (11) for nonconducting domain $\Omega_n$ which are used together with conditions (21) and (22) preliminary expressed in a weak form. This yields the system of coupled integral equations for mixed vector-scalar formulation as follows

$$\int_{\Omega_c} \left(1/\mu_c \, \boldsymbol{\nabla} \times \boldsymbol{A}\right) \cdot (\boldsymbol{\nabla} \times \boldsymbol{w}) dv - \int_{\Gamma_i} (\mu_n \boldsymbol{n} \times \boldsymbol{\nabla} V_m) \cdot \boldsymbol{w} ds = \int_{\Omega_c} \boldsymbol{j} \cdot \boldsymbol{w} \, dv + \int_{\Omega_c} (\boldsymbol{\nabla} \psi) \cdot \boldsymbol{w} \, dv, \qquad (23)$$

$$\int_{\Omega_n} (\mu \boldsymbol{\nabla} V_m) \cdot (\boldsymbol{\nabla} \zeta) dv + \int_{\Gamma_i} \boldsymbol{n} \cdot (\boldsymbol{\nabla} \times \boldsymbol{A}) \zeta ds - \int_{\Gamma_{cut}} \mu \zeta \boldsymbol{n} \cdot (\boldsymbol{\nabla} V_m^+ - \boldsymbol{\nabla} V_m^-) ds = 0, \qquad (24)$$

$$\int_{\Omega_c} \boldsymbol{A} \cdot (\boldsymbol{\nabla} \zeta) dv = 0, \qquad (25)$$

Some analytical results regarding to ellipticity of the bilinear form as well as the uniqueness of solution for such type of systems can be found elsewhere [20-23]. Next steps toward finding the numerical solutions such as meshing the computational domain with small nonoverlapping elements, approximating the unknown vector and scalar potentials on a finite-element mesh by the local edge and nodal basis functions, respectively, as well as discretizing the weak forms of PDEs to build up the local matrices, assembling the local matrices to a global matrix and, finally, solving linear algebraic systems are performed with COMSOL Multiphysics software specially adjusted to obtain the stable solutions.

It should be noted that mixed vector-scalar formulation has been already implemented in the *Rotating Machinery, Magnetic* interface of COMSOL Multiphysics, where it is specifically designed for modeling the magnetic rotating machines [25]. In particular, the interface uses the moving mesh approach to enable the relative motion of stators and rotors. To be meshed separately, these two parts must be treated as separated geometrical objects combined into an assembly pair with the prescribed rotor-stator coupling. Although it is possible, the application of this interface for modeling of stationary magnets would lead however to the unnecessary complexity.

In this paper, we employ instead the two other interfaces of this software, the *Magnetic Fields* and the *Magnetic Fields, No Currents* usually used separately to solve either for magnetic vector, or for total scalar potential. We implement the mixed formulation by manually coupling these interfaces to each other along their common interfacing boundary to solve for combination of both potentials.

2. Modeling Examples

    a. Models and modeling approaches

In order to evaluate the numerical potential of the mixed vector-scalar formulation against the MVP formulation, we use two magnetization models, the Helmholtz coil surrounded by air and the dipole magnet shown in Fig.1.

**Helmholtz coil carrying DC current:** The model consists of the two circular coils arranged in parallel at equal distances above and below the median plane of the large spherical air domain used for magnetic insulation. There are no nonlinear materials in this model.

**Dipole magnet:** The model consists of the two poles arranged in parallel at equal distances above and below the median plane of the large spherical air domain used for magnetic insulation. Either pole of the magnet includes the circular coil driven by the DC current as well as the yoke, four spiral sectors and other constituents made of the nonlinear ferromagnetic magnetic materials.

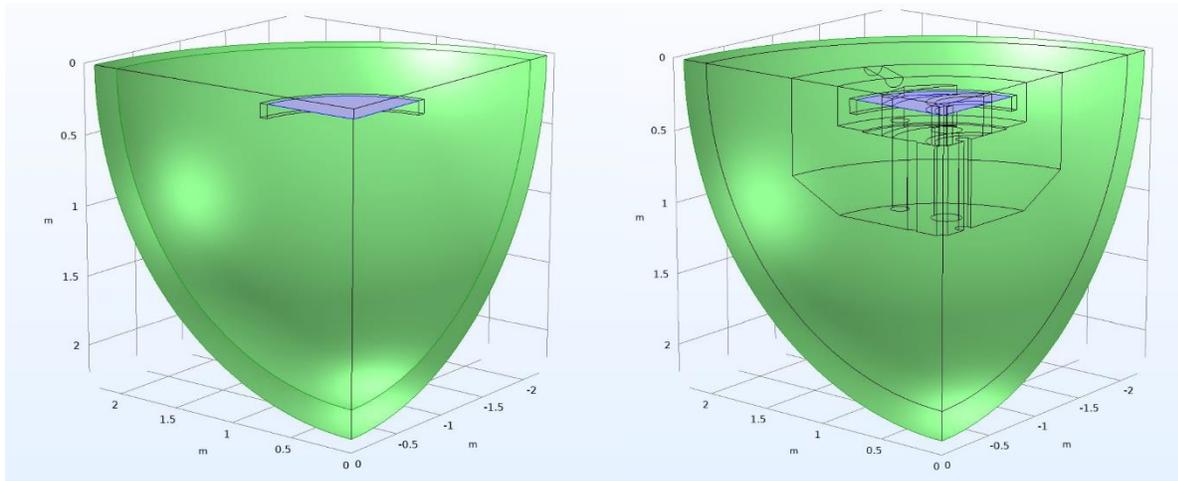

Figure 1: The 1/8-th parts of the model geometries of Helmholtz coil (left) and dipole magnet (right) surrounded by the insulation air domain with constructed cut surfaces.

The geometry of circular coil is identical for both models. It represents the axisymmetric hollow cylinder formed by rotation around the $z$-axis of a rectangle lying in $(z, x)$-plane. The coil is of the multiturn type and therefore modeled as a homogenized current carrying region with multiple wires arranged and placed in a potting material. The excitation DC current of the same value for both models is applied to the coil internal cross-section boundary. A coil numeric analysis is made prior the field calculation to compute the magnitude and direction of the current flow in conductor. The coil current densities are shown in Fig.2.

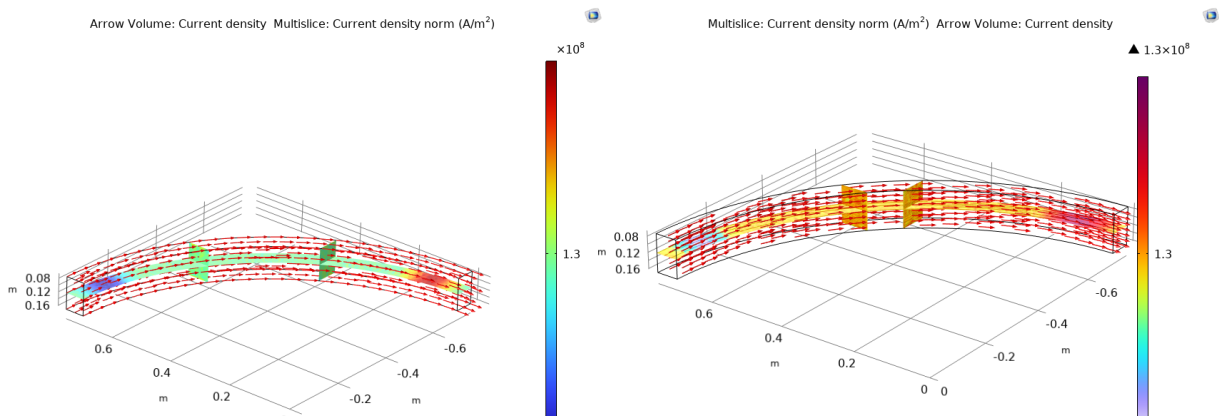

Figure 2: The applied coil current densities in the models of Helmholtz coil (left) and dipole magnet (right).

The geometry of insulating air domain is also identical for both models. It represents a large sphere surrounded by the boundary layer, whose thickness is scaled towards infinity to mimic virtually the infinite element domain. Additionally, a single cut surface is constructed and added to geometries of both models when the mixed vector-scalar formulation is applied. Since the size and position of this surface do not affect the physics, the minimal area criterium is used for optimal construction providing the lower increase of the total number of DOFs after nodal doubling on the cut surface.

Both models possess three planes of symmetry allowing to truncate the whole problem domain and reduce the computational cost by exploring the 1/8-th part of the model geometry and obtaining the results for the whole model. The mirror symmetry allows to cut the model geometries along the median plane and impose either the Perfect Magnetic Conductor boundary condition (MVP formulation), or the Zero Magnetic Scalar Potential boundary condition (mixed formulation), whereas the four-fold axial symmetry is used to cut the remaining parts of the model geometries along the $(z, x)$ and $(z, y)$-planes and impose the Magnetic Insulation boundary conditions. The role of these boundary conditions is to mimic the entire geometry while exploiting its 1/8-th part.

In the MVP formulation, the magnetic vector potential is applied to the whole computational domain of both models. However, in the mixed formulation, the vector potential is only applied to the region of coil while the air and other nonconducting regions are modeled with magnetic scalar potential. Both potentials are then coupled along the coil boundary by applying the interface conditions (21) and (22) to ensure field continuity.

The two studies relied on the same model geometry of both examples are performed and compared. The first focuses on the numerical performance of the mixed vector-scalar formulation. The second repeats a similar simulation by using the magnetic vector potential formulation. The main quantities of interest are the field distributions over the median plane as well as along the radial and azimuthal directions in the aperture area. For FEM analysis of each model, the conforming mesh is generated with almost the same number of finite elements for different potential formulations despite the additional cut surface is constructed and added to geometries of both models when the mixed formulation is used. The minimal mesh quality is optimized to ensure the convergence and stability of solutions. The edge and Lagrange shape functions up to third order are used for approximation of magnetic vector and scalar potentials, respectively. The mixed vector-scalar formulation is solved in fully coupled approach with the same for both potentials direct PARDISO solver based on multifrontal factorization of the stiffness matrix to ensure convergence of solution. For purpose of comparison, the same solver is then employed for the MVP formulation of both models. The gauge fixing for $\boldsymbol{A}$-field is used with the direct solver as the necessary condition to ensure the convergence of solution. Its side-impact to the convergence of nonlinear solver in the presence of the nonlinear materials for model of dipole magnet is overcome by splitting the solution process into several steps, each with its own behavior of the permeability specified as a constant, linear, and nonlinear, respectively, and with the initial conditions resulting from simulation of the previous step.

In the mixed formulation, the magnetic field outside the coil region of both models is obtained by solving first for the magnetic scalar potential and taking then the negative gradient from the solution result. For this reason, the potential discontinuities across the cut surfaces do not violate the continuity of the magnetic field as demonstrated in Fig.3.

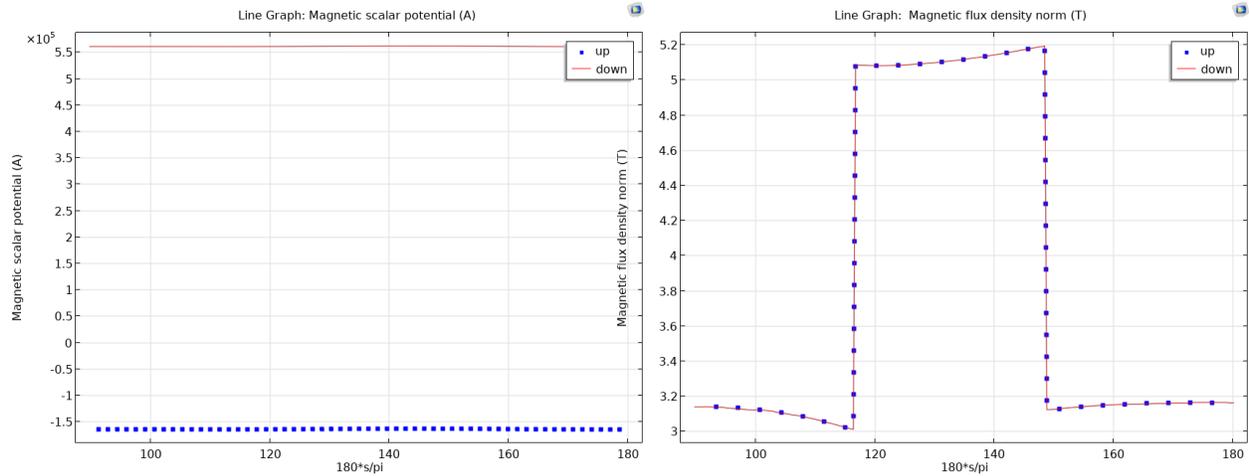

Figure 3: Azimuthal distributions of the magnetic scalar potential (left) and the flux density norm (right) along up and down sides of the cut plane. The jump of the potential is exactly equal to the value of applied coil current.

b. Helmholtz coil

The simulation results obtained for finite-element modeling of Helmholtz coil with the use of COMSOL software are shown in figures 4 and 5 and summarized in Table 1.

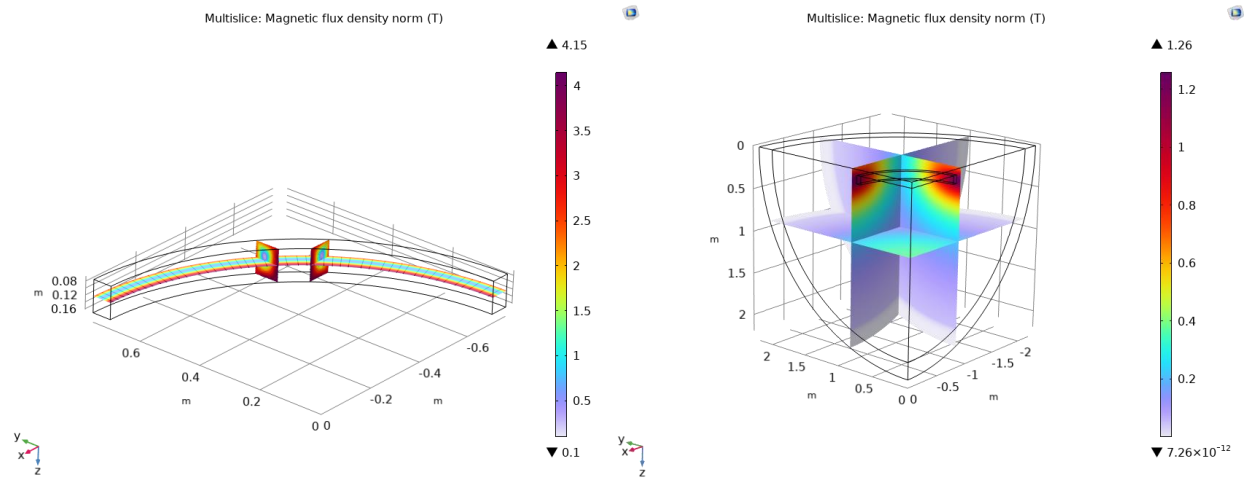

Figure 4: Distributions of the magnetic flux density norm inside the coil (left) and the air (right) regions.

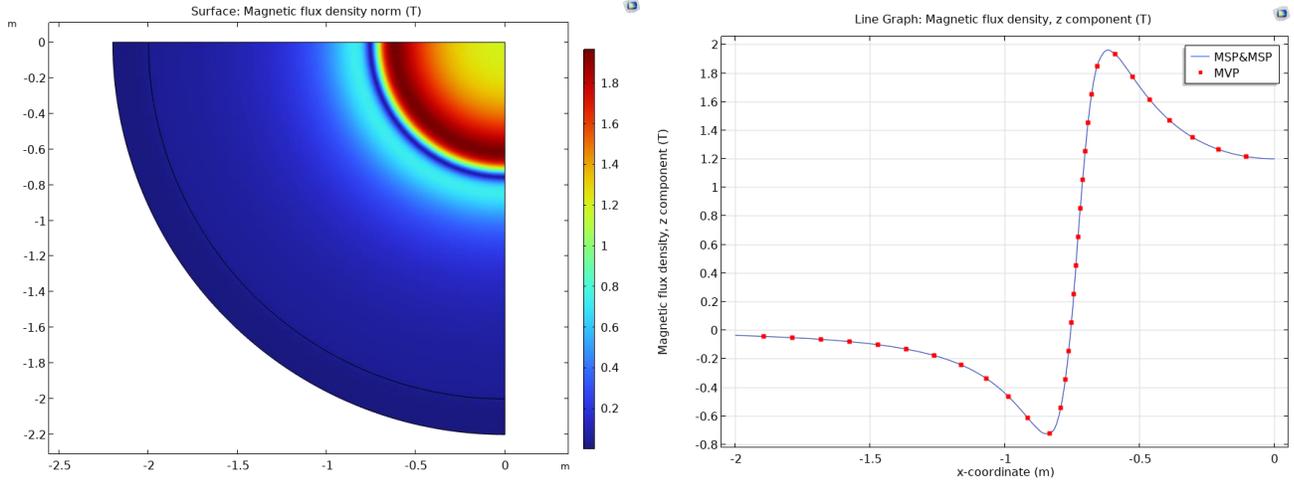

Figure 5: Distributions of the magnetic flux density norm over the median plane (left) and the z-component of the magnetic flux density along the radial direction (right). Solid and points refer to calculations with MVP&MSP and MVP formulations, respectively.

Both formulations produce the excellent results in qualitative and quantitative agreement. As compared to MVP formulation, the mixed formulation requires much less computational resources for finite-element modeling of Helmholtz coil. The reduction of DOVs amounts to a factor of 4.2, the RAM to a factor of 6, and the time of computation to a factor of 16.9 for relative error between the two formulations of 0.0001 Tesla, or 1 Gauss.

Table 1: Summary of comparison between formulations used for Helmholtz coil

|         | Element order | Number of FEs | Number of DOVs | Memory (GB) Phys/Virtual | Time of computation |
|---------|---------------|---------------|----------------|--------------------------|---------------------|
| MVP&MSP | 3/3           | 48 509        | 295 993        | 7.54/22.5                | 9s                  |
| MVP     | 3             | 46 480        | 1 256 270      | 45.29/63.69              | 2m 32s              |

c. Dipole magnet

Figures 6 and 7 present the simulation results obtained for finite-element modeling of dipole magnet with the use of COMSOL software and summarized in Table 2. Both formulations produce the expected results in excellent qualitative and quantitative agreement. As compared to MVP formulation, the mixed formulation requires much less computational resources for finite-element modeling of dipole magnet. The reduction of DOVs amounts to a factor of 4.85, the RAM to a factor of 10, and the time of computation to a factor of 29.6 for relative error between the two formulation of 0.0003 Tesla, or 3 Gausses.

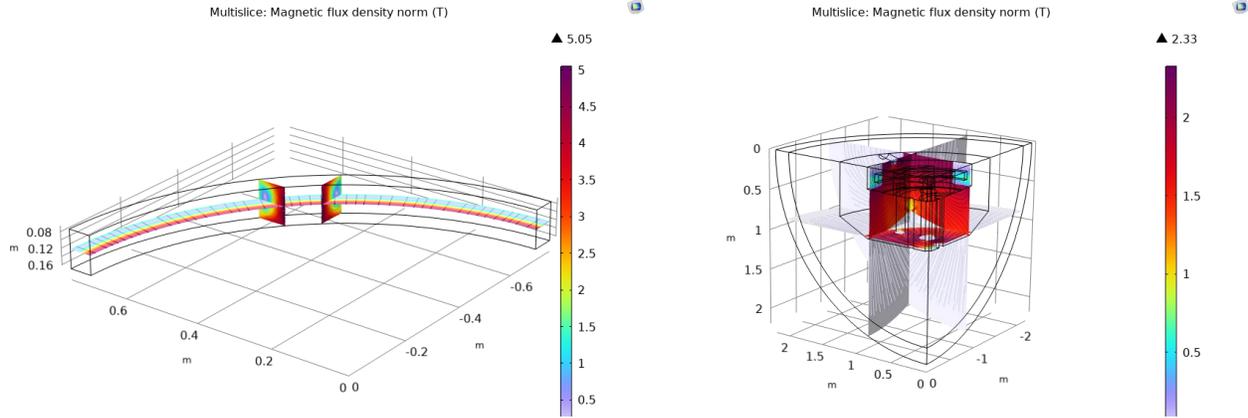

Figure 6: Distributions of the magnetic flux density norm inside the coil (left) and the pole (right) regions.

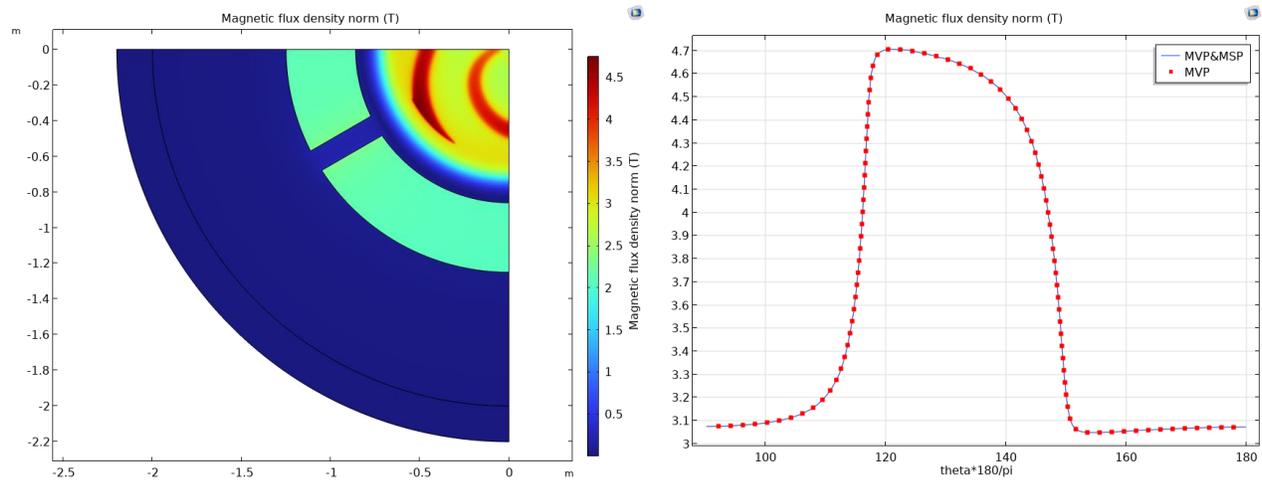

Figure 7: Distributions of the magnetic flux density norm over the median plane (left) and along the azimuthal direction (right). Solid and points refer to calculations with MVP&MSP and MVP formulations, respectively.

Table 2: Summary of comparison between formulations used for dipole magnet

|  | Element order | Number of FEs | Number of DOVs | Memory (Gb) Phys/Virtual | Time of computation | Number of iterations |
|---|---|---|---|---|---|---|
| MVP&MSP | 3/3 | 423 198 | 2 049 396 | 38.19/56.59 | 8m 33s | 7 |
| MVP | 3 | 414 840 | 9 958 301 | 393.15/443.24 | 4h 12m 46s | 8 |

Conclusion

In this paper, we propose to use for finite-element modeling of the large-scale magnetization problems in the presence of applied currents and highly nonlinear materials the combination of magnetic vector and total scalar potentials as an alternative to the well-known for its quality of calculation but computationally expensive vector potential formulation. The potentials are applied to conducting and nonconducting parts of the problem domain, respectively and coupled together across their common interfacing boundary. For nonconducting regions of the problem domain, the thin cuts are constructed to ensure their simply connectedness, and therefore the consistency of the mixed formulation. The numerical performance of finite-element modeling in terms of combined potentials is assessed against the magnetic vector potential formulation for two magnetization models, the Helmholtz coil, and the dipole magnet. We show that mixed formulation can provide a substantial reduction in the computational cost as compared to its vector counterpart for a similar accuracy of both methods.

Acknowledgements

The computational support from HybriLIT Heterogeneous Computing Platform (LIT, JINR) is acknowledged.